\documentclass[twocolumn,prl]{revtex4}
\begin{document}

\title{Comment on  ``Universal Thermal Radiation Drag on Neutral Objects''}
\author{P. A. Maia Neto and C. Farina}
\affiliation{
Instituto de F{\'\i}sica, UFRJ, Caixa Postal
68528, Rio de Janeiro RJ, 21945-970, Brasil}
\date{\today}

\maketitle

In a recent letter~\cite{letter}, Mkrtchian and co-workers 
calculated the radiation pressure force on a moving body, assuming the 
electromagnetic field to be at temperature  $T,$ and the velocity to be much smaller than $c.$ 
They considered both dielectrics and conductors, and related 
the effect to Casimir dissipative forces. 
Here we claim that their approach may only apply in the 
Rayleigh-Ganz scattering regime, which corresponds to
very small particles and/or electromagnetically rarefied media~\cite{Hulst}. 
 Moreover, we argue
that their interpretation in terms of the Casimir effect is misleading, since 
vacuum fluctuations do not contribute in the (implicitly assumed) regime of 
uniform motion.   

The polarization at a given point of a spatially local medium  depends on the 
total electric field at this point, which  is
a superposition of the incident electric field and 
the electric field produced by the polarized body. 
However, when computing the force on the moving body, the authors 
neglect the field produced by  the medium 
and only take into account the incident field, corresponding to the 
freely propagating field whose correlation function is given by their equation (5). 
This might be consistent if the typical size $a$ of the body is much smaller than the 
 field wavelengths both outside  and inside the medium, so that it effectively behaves as a 
single electric dipole (Rayleigh regime): $a\ll \lambda$ and $a\ll \lambda/|\hat{n}|,$ 
where $\hat{n}$ is the complex refractive index of the medium~\cite{foot}. 
Even if the body is not very small, the electric field produced by the medium may be neglected when 
computing the polarization if  $\hat{n}$ is sufficiently close to one: 
$2\pi a|\hat{n}-1|/\lambda\ll 1.$
This less restrictive condition defines the Rayleigh-Ganz scattering regime~\cite{Hulst}. 

Once the range of validity of the authors' approach is understood, 
some intriguing features of
their results are (at least partially) clarified. 
It is known that non-absorbing 
objects moving in a thermal field also experience the same type of 
drag force~\cite{jr}\cite{PRA2002}, whose physical 
interpretation relies on the
Doppler shift of the thermal photons upon reflection (or more generally scattering)
by the moving body~\cite{jr}.  
However, the authors' expression for the force, as given by their equation (12), vanishes  
in the case of zero absorption. 
Hence the derivation of the force on a non-absorbing small body
requires a less crude approximation,  involving a higher
power of $a/\lambda,$ just as in the derivation of the extinction cross section from the 
optical theorem~\cite{Hulst}.  

For conductors, 
the force, as given by equations (16) and (17) of \cite{letter},
is  proportional to the volume of the moving body.
This only holds if the body is much smaller than the skin depth for penetration of the 
field inside the 
conductor (so as to be consistent 
with the Rayleigh approximation which requires a uniform field throughout the body), 
a condition that follows from the inequality $a\ll \lambda/|\hat{n}|$
mentioned above. In the opposite limit, the force must  be proportional to the
cross section area of the body.
This is of course always the case for perfect reflectors~\cite{PRA2002} (see particularly 
\cite{deco} for the force on a moving perfectly-reflecting sphere). 
These remarks explain why the authors' result has no finite value in the limit of infinite 
conductivity, since this limit is not consistent with the approximations employed 
throughout the paper. 

Despite of the authors' comments on the connection with the Casimir effect, their result for the force
does not contain any contribution of vacuum fluctuations, so that it  
vanishes at zero temperature, as can be checked by taking the limit
$\beta\rightarrow\infty$ in equation (12). 
This is a consequence of the implicit assumption of constant velocity 
when deriving the field correlation function
in the co-moving frame, as given by equations (5) and (6). 
On the other hand, zero-temperature dissipative forces may appear for a {\it single}
moving body provided that the motion is non-uniform~\cite{PRA2002}\cite{dissipation}, 
an effect closely related to the fluctuations 
of the Casimir force on the body~\cite{fluctuations}.

The authors' comment on the connection with the Casimir effect is also misleading for 
a second reason: boundary effects play no role in  Rayleigh scattering. 
Hence geometrically determined resonances (like Mie resonances)
cannot have any effect on their results, in contradiction with their
final statement.


\begin{thebibliography}{99}


\bibitem{letter} V. Mkrtchian, V. A. Parsegian, R. Podgornik, and W. M. Saslow, 
Phys. Rev. Lett. {\bf 91}, 220801 (2003). 

\bibitem{Hulst} H. C. van de Hulst, {\it Light Scattering by Small Particles}, 
 (Dover, New York, NY, 1981).

\bibitem{foot} For a thermal field, relevant wavelengths are of the order 
$\lambda\sim 2\pi \hbar c/(k_B T),$ where $k_B$ is the Boltzmann constant.  

\bibitem{jr}  M. T. Jaekel and S. Reynaud, Phys. Lett. A {\bf 172}, 319 (1993).

\bibitem{PRA2002} L. A. S. Machado, P. A. Maia Neto and C. Farina, Phys. Rev. D {\bf 66}, 105016 (2002). 

\bibitem{deco} S. Reynaud, P. A. Maia Neto, A. Lambrecht, and 
M.-T. Jaekel, Europhys. Lett. {\bf 54}, 135 (2001).

\bibitem{dissipation} S. A. Fulling and P. C. W. Davies, Proc. R. Soc. London
 \textbf{A348}, 393 (1976);
L. H. Ford and A. Vilenkin, Phys. Rev. D \textbf{25}, 2569
(1982).

\bibitem{fluctuations} P. A. Maia Neto and S. Reynaud, Phys. Rev. A {\bf 47}, 1639 (1993).


%


\end{thebibliography}
\end{document}